\documentclass[aps,prl,reprint,groupedaddress]{revtex4-1}
\usepackage{graphicx, color, soul}
\usepackage{siunitx}
\usepackage{amsmath}

\begin{document}

\title{Fractal waveguide arrays induce maximal localization}

\author{Jonathan Guglielmon$^{1}$ and Mikael C. Rechtsman$^{1}$}
\affiliation{$^{1}$Department of Physics, The Pennsylvania State University, University Park, PA 16802, USA}

\date{\today}

\begin{abstract}
The ability to transmit light through an array of closely packed waveguides while minimizing interwaveguide coupling has important implications for fields such as discrete imaging and telecommunications. Proposals for achieving these effects have leveraged phenomena ranging from Floquet-induced flat bands to Anderson localization. Here we demonstrate that, for strongly detuned waveguides, optimal localization is achieved not by disorder but by fractal potentials. We further show that, in both 1D and 2D, these structures possess a localization-delocalization phase transition.
\end{abstract}

\maketitle

The transmission of light through a discrete array of waveguides has important applications ranging from imaging to telecommunications \cite{hopkins1954,Richardson2013,Winzer2014}. In these contexts, each waveguide serves as an independent channel through which image data or other information is transmitted. A fundamental and ubiquitous problem that arises in these systems is the presence of interwaveguide crosstalk, which originates from evanescent coupling between neighboring waveguides. A variety of proposals have been put forward to circumvent this problem. One potential solution utilizes dynamic localization \cite{Eisenberg2000,Longhi2006,Szameit2010,Garanovich2012,Crespi2013}, where a periodic spatial modulation of the waveguide trajectory causes renormalized nearest-neighbor couplings to vanish. This is equivalent to introducing a Floquet drive that is then tuned to generate a flat quasienergy band. Alternatively, one can take the waveguides to be straight and either arrange them into a structure, such as a Lieb or Kagome lattice, that exhibits a non-dispersive flat band \cite{Vicencio2014,Silva2014,Mukherjee2015}, or introduce a detuning between adjacent waveguides by modifying their propagation constants so as to reduce the effective strength of the interwaveguide coupling \cite{Koshiba2009}. Here, the waveguide propagation constants are tuned either by adjusting the waveguide diameter or by varying the refractive index of the core. Another avenue takes inspiration from the phenomenon of Anderson localization \cite{AndLoc1, AndLoc5,Mafi1,Mafi2,Mafi3} and uses disorder to localize the system's eigenstates, thereby suppressing diffraction. 

Each of these approaches comes with a distinctive set of limitations. Structures containing flat bands, such as the Lieb lattice, also contain additional bands that are dispersive. Since an arbitrary input state will excite a combination of the flat and dispersive bands, general states are not preserved under evolution through the structure. Additionally, higher-neighbor couplings add dispersion to the bands so that coupling is only suppressed out to a finite propagation distance determined by the associated coupling length. In contrast, disorder localizes every eigenstate and this localization persists out to infinite propagation distance. However, the degree to which crosstalk is reduced crucially depends on the localization length which, in two dimensions, can be large. It is therefore natural to pose the question: to what extent can the localization lengths be reduced by more precisely tailoring  the configuration of individual waveguide propagation constants?

In this work, we demonstrate that, in the regime in which waveguide propagation constants can be strongly detuned, optimal localization is achieved not by disorder but by self-similar configurations in which sites are recursively detuned from one another over increasingly larger spatial scales. Being aperiodic, these structures possess the advantages of disorder arising from eigenstate localization, but they surpass disorder in their ability to tightly confine the eigenstates. We characterize these structures, noting the existence of a localization-delocalization phase transition in both 1D and 2D. We also characterize a related series of periodic lattices with increasingly larger unit cells that are able to suppress diffraction out to increasingly larger propagation distances. 

\textit{Results}.\textemdash With imaging applications in mind, we consider light evolving through an array of coupled waveguides and note that, while we develop our results in the photonic context, they in fact generalize to systems that can be modeled as a lattice of coupled degrees of freedom with a tunable onsite potential. Such coupled waveguide arrays are governed by the paraxial Schr{\"o}dinger equation\textemdash an equation identical to the Schr{\"o}dinger equation for a quantum mechanical particle except that the time coordinate is replaced by the propagation distance $z$ measured along the axial direction of the waveguides. In such a system, the waveguides act as lattice sites that possess bound modes whose profiles weakly overlap with their neighbors so that the system can be modeled using tight-binding theory. The onsite energy, $V_i$, of each waveguide is determined by the waveguide propagation constant, which can be controlled by varying either the waveguide diameter or the refractive index of the waveguide core. This yields a tight-binding Schr{\"o}dinger equation
\begin{equation}\label{eqn_tight_binding}
i \partial_z \psi_i(z) = \sum_{ j} (c_{ij} + V_i\delta_{ij})\, \psi_j(z)
\end{equation}
where $\psi_i$ is the overlap of the electric field profile with the bound mode of the $i$-th waveguide and $c_{ij}$ is the coupling constant between sites $i$ and $j$. 

To send an image through such a structure, one injects an electric field profile $\psi_i$ at the input facet with intensity amplitudes $|\psi_i|^2$ equal to the image intensity at a pixel situated at the location of the waveguide. The light propagates through the structure and the corresponding intensity amplitudes $|\psi_i'|^2$ observed at the output facet produce the transmitted image. In this context, eigenstate localization is desirable in order to keep $|\psi_i'|^2$  as similar as possible to $|\psi_i|^2$. Motivated by the ability of disorder to induce localization, we pose the question: what is the optimal way to distribute the onsite energies, $V_i$, so as to produce maximal localization?

To answer this question, we define an objective function, $S(z)$, that measures how much the intensity distribution of input states will change during evolution over a propagation distance $z$. In choosing such a function, we note that the sum of squared differences in position space intensities between the initial and final states is given by
\begin{equation}
\sum_i \left(\left|\psi_i\right|^2-\left|\psi_i'\right|^2\right)^2 = \sum_i \Big(\langle \psi| M_i |\psi\rangle\Big)^2
\end{equation}
where the $M_i$ are operators defined by $M_i = |e_i\rangle \langle e_i| - |u_i\rangle \langle u_i|$. Here $|e_i\rangle$ is the unit vector with all except the $i^\text{th}$ entry equal to zero and $\langle u_i|$ is the $i^\text{th}$ row vector of the operator $U(z)$. Keeping $|\psi\rangle$ general, we then seek to minimize $\sum_i ||M_i||^2$, where we use the entry-wise matrix norm $||A||^2 = \sum_{ij} |A_{ij}|^2$. It is simple to show that this is equivalent to maximizing $\sum_i |\langle u_i|e_i\rangle|^2=\text{tr} \left|U(z)\right|^2$, where the absolute value is taken element-wise when $U(z)$ is expressed in the position basis. In general, $\text{tr} \left|U(z)\right|^2$ can oscillate rapidly as a function of $z$, so that a solution that is optimal at $z$ may fail to be optimal at a nearby point $z + \delta z$. To avoid this problem, we perform an integral over $z$ so as to capture information about the behavior over the entire interval:
\begin{equation} \label{eqn_objfunc}
S(z) = \frac{1}{nz}\int_0^zdz'\,\text{tr} \left|U(z')\right|^2.
\end{equation}
Here we have further normalized the function by the number of sites, $n$, so that $S(z) \in [0,1]$ with $S(z) = 1$ implying that an arbitrary single-site injection will remain perfectly localized at the injection site out to propagation distance $z$.

\begin{figure}
\centering
\includegraphics[width =1\columnwidth]{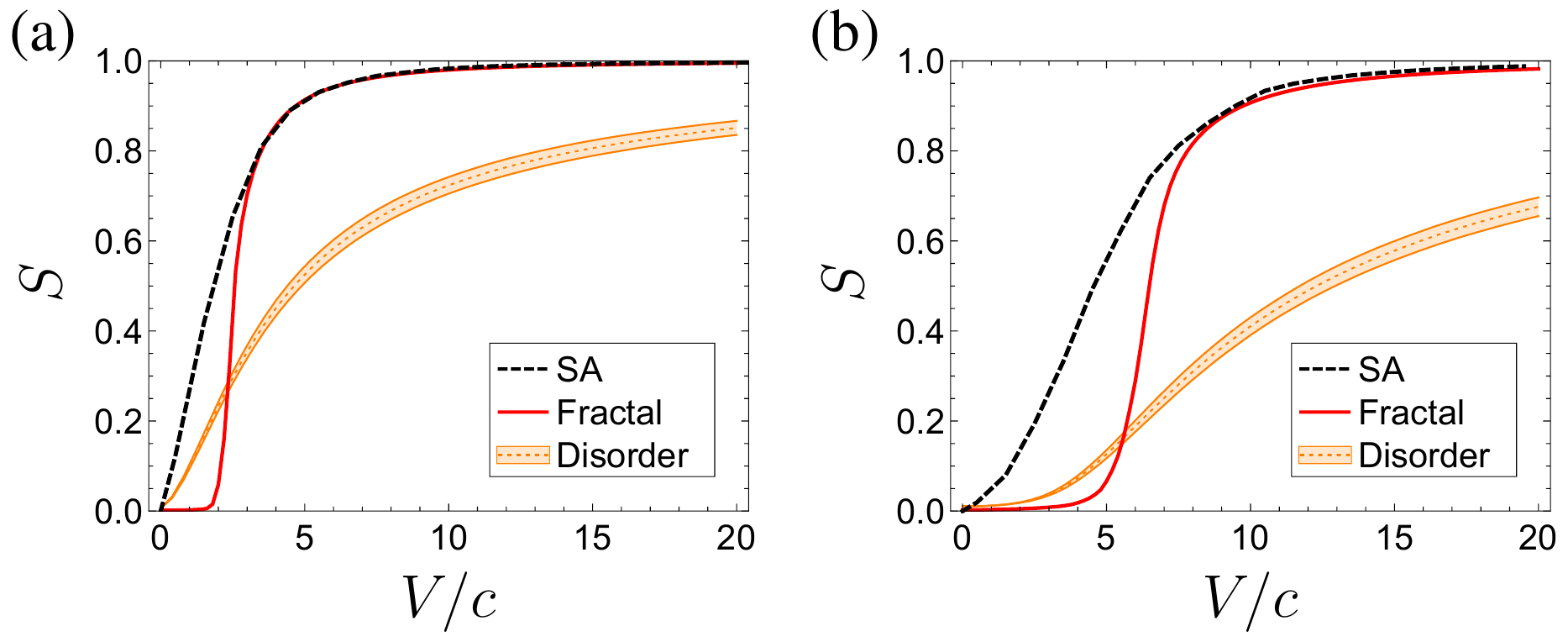}
\caption{\label{fig_comparison} Objective function for different onsite energy configurations. Panels (a) and (b) respectively show the results for 1D and 2D systems. The dashed black curve was obtained using simulated annealing to optimize the onsite energies. The dashed orange curve shows the results for onsite disorder averaged over many disorder realizations and the shaded orange region corresponds to one standard deviation from the mean. The red curve shows the results for the fractal structure described in the text, using $\alpha = 0.25$ and $\alpha = 0.36$ for the 1D and 2D cases, respectively. In the strongly detuned regime, the fractal structure outperforms disorder in both 1D and 2D and becomes nearly optimal.}
\end{figure}

In our optimization, we will take the infinite propagation distance limit of $S(z)$ and only evaluate $S(z)$ at finite propagation distance when  characterizing a set of periodic structures that are closely related the self-similar potential that we will introduce later. In this limit, $S(z)$ bears a close relation to the participation ratios of the energy eigenstates. In particular, assuming the system possesses no degeneracies, we have 
\begin{equation}\label{eqn_infz}
\lim_{z\to \infty} S(z) = \left\langle P_E\right\rangle
\end{equation}
where $\left\langle P_E\right\rangle$ represents the average of the participation ratios over all the energy eigenstates (see Supplemental Material \cite{suppmat} for details).

In this work, we will restrict attention to potentials defined over an equally spaced linear array in 1D and a square lattice in 2D. We note, however, that the structures we uncover are ultimately applicable to a more general class of $n$-partite lattices satisfying the property that the $n$ independent sublattices form rescaled copies of the original lattice. This includes the important case of the triangular lattice, which is used in applications that seek to maximize the density of lattice sites. We perform an optimization of $S(z)$ at $z\to \infty$ (i.e., Equation (\ref{eqn_infz})) using simulated annealing on a 256 site chain in 1D  and a $16\times16$ lattice in 2D. In both cases, we impose periodic boundary conditions and include only nearest-neighbor coupling $c$. We restrict the onsite energies to lie within a finite interval $[-V,V]$ and perform independent optimizations for different values of $V$.

\begin{figure}
\centering
\includegraphics[width = 1\columnwidth]{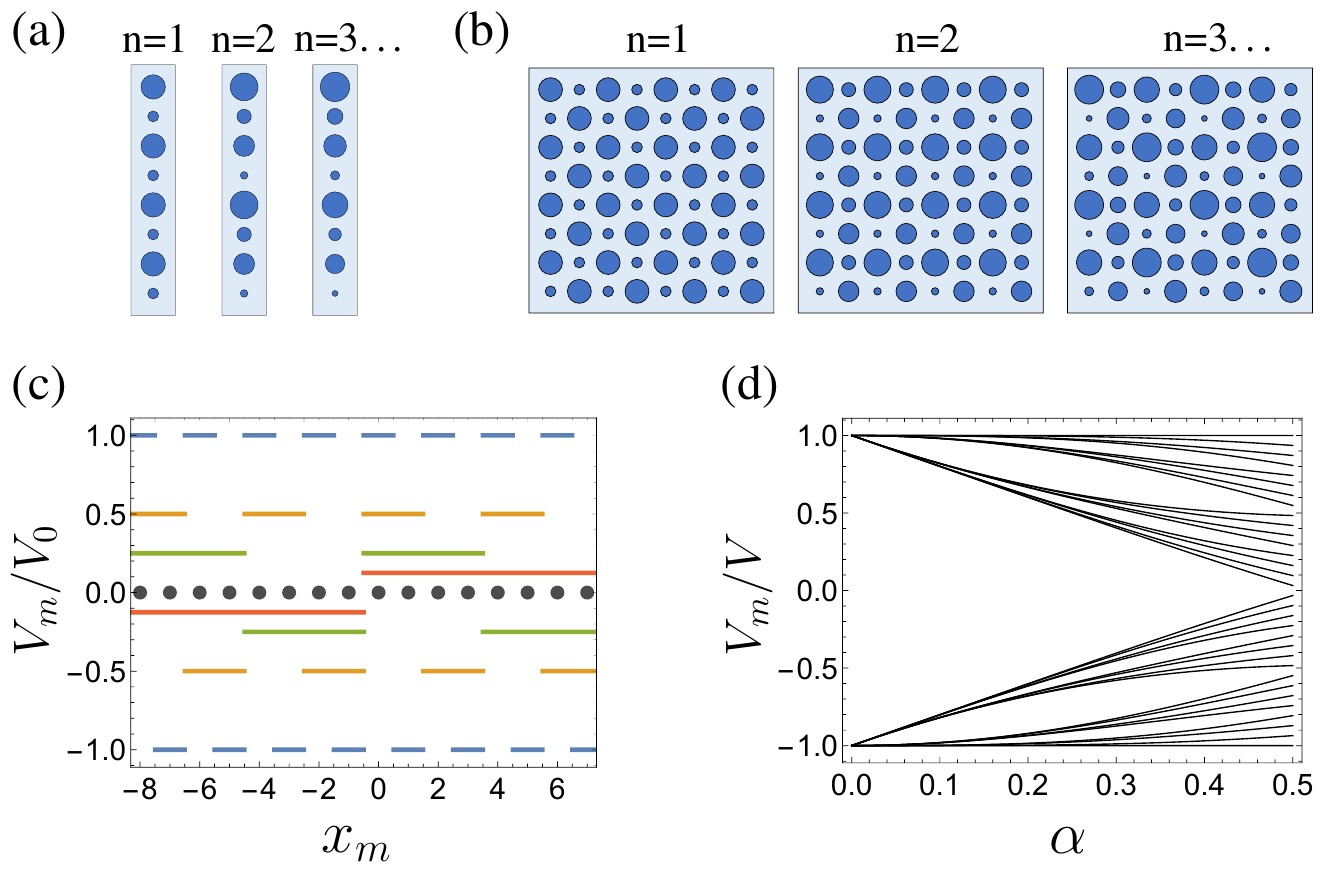}
\caption{\label{fig_structure}Construction of the self-similar potentials described in the text. Panel (a) shows the construction of the 1D case and panel (b) the 2D case, where the onsite energies have been encoded in the radius of the circles. The structure is generated by recursive application of a detuning procedure in which each increase in the recursion level, $n$, results in an exponentially larger unit cell, with the full structure obtained in the $n\to \infty$ limit. Panel (c) shows the decomposition of the 1D potential into square waves, with the sites of the lattice overlaid on the $x$-axis. Panel (d) shows the dependence of the onsite energies on the parameter $\alpha$ for a 32-site chain. Onsite energies at a given $\alpha$ are simply sorted by energy and do not directly correspond to locations on the lattice.}
\end{figure}

The results of the optimization are plotted in Figure \ref{fig_comparison} and provide a baseline standard for ideal performance. For comparison, we have also shown the corresponding curve for a disordered potential using the same system size as in the optimization and averaged over $1000$  disorder realizations. We note, in particular, that the optimum is achieved not by disorder but by some other potential. However, as the optimizations are performed for a finite system size, it is desirable to extract structures that perform at the level indicated by the optimization and that are amenable to a precise description that enables us construct the structure for arbitrarily large system size so as to verify that the optimized results do not rely on finite size effects. 
We find that this can be done in the strong potential limit, where optimal performance can be achieved by a self-similar potential that we construct below. While this highly ordered potential performs optimally, we note that our optimization suggests that the optimum may  be non-unique or nearly degenerate with other structures. In particular, the numerically optimized structures themselves both contain features associated with the potential that we construct below, as well as features that differ from this potential. The fact that these features do not significantly change the performance of the structures suggests the non-uniqueness of the optimum.

We now demonstrate that the degree of localization achieved by the numerically optimized structures can be achieved by a self-similar fractal potential that can be constructed analytically. We begin by fixing an onsite energy scale $V_0$ and a dimensionless parameter $\alpha \in [0,1/2]$ that controls the structure of a Cantor set from which the onsite energies are sampled (see Supplemental Material \cite{suppmat}). Noting that the lattice (a 1D linear array or 2D square lattice) can be made to be bipartite, we subdivide the lattice into two sublattices to which we respectively assign onsite energies of $\pm  V_0$. Since each of the sublattices themselves form rescaled copies of the original lattice (rotated by $\pi/4$ in the 2D case), we can independently repeat the detuning procedure on each sublattice, this time using a smaller detuning of $\pm \alpha  V_0$. In general, at a recursion level $n$, we apply a detuning of $\alpha^{n}  V_0$ to obtain the structure at level $n+1$. This procedure is illustrated in Figure \ref{fig_structure}. When this process is iterated sending $n\to \infty$, the structure converges to an aperiodic potential with onsite energies contained in the interval $[-V,V]$, where $ V = V_0/(1-\alpha) $.  In the 1D case, the resulting potential can be written as:
\begin{equation}\label{eqn_frac1d}
V_m =V_0\sum_{k=0}^\infty\alpha^k s\left(\frac{m+1/2}{2^{k+1}}\right)
\end{equation}
where $s(x)$ is an odd, unit-period square wave that alternates between $\pm 1$. Here the potential has been evaluated at site $x_m = ma$, where $a$ is the lattice constant. See Figure \ref{fig_structure}(c) for an illustration of this square wave expansion. In the 2D case, the potential takes on a similar form:
\begin{equation}
V_{mn} = V_0 \sum_{k=0}^\infty \left[ \alpha^{2k} s\left(\frac{m+1/2}{2^{k+1}}\right) + \alpha^{2k+1}s\left(\frac{n+1/2}{2^{k+1}}\right)\right]
\end{equation}
where the pair of indices on $V_{mn}$ indicate that the potential has been evaluated at site $\mathbf{x}_{mn} = m\mathbf{R}_1 + n\mathbf{R}_2$ with $\mathbf{R}_1=a(1,0)$ and $\mathbf{R}_2=a(1,1)$. 

In Figure \ref{fig_comparison}, we have plotted the objective function for these structures to show that, in the strong potential limit, they perform optimally in comparison to the results from simulated annealing. We note that we have increased the system size in producing the plots for these structures to a $1024$ site chain in 1D and a $32\times32$ structure in 2D to demonstrate that our fractal model achieves optimal performance for system sizes beyond that used in the optimization. We have also included a set of animations in the Supplemental Material \cite{suppmat} that show the output obtained by evolving an image through the structure, with the results for the fractal and disordered potentials compared side-by-side. We have included separate animations for the cases of coherent imaging and incoherent imaging \cite{goodman2017}, with the fractal yielding noticeable improvements over disorder in both cases.

\begin{figure}
\centering
\includegraphics[width = 1\columnwidth]{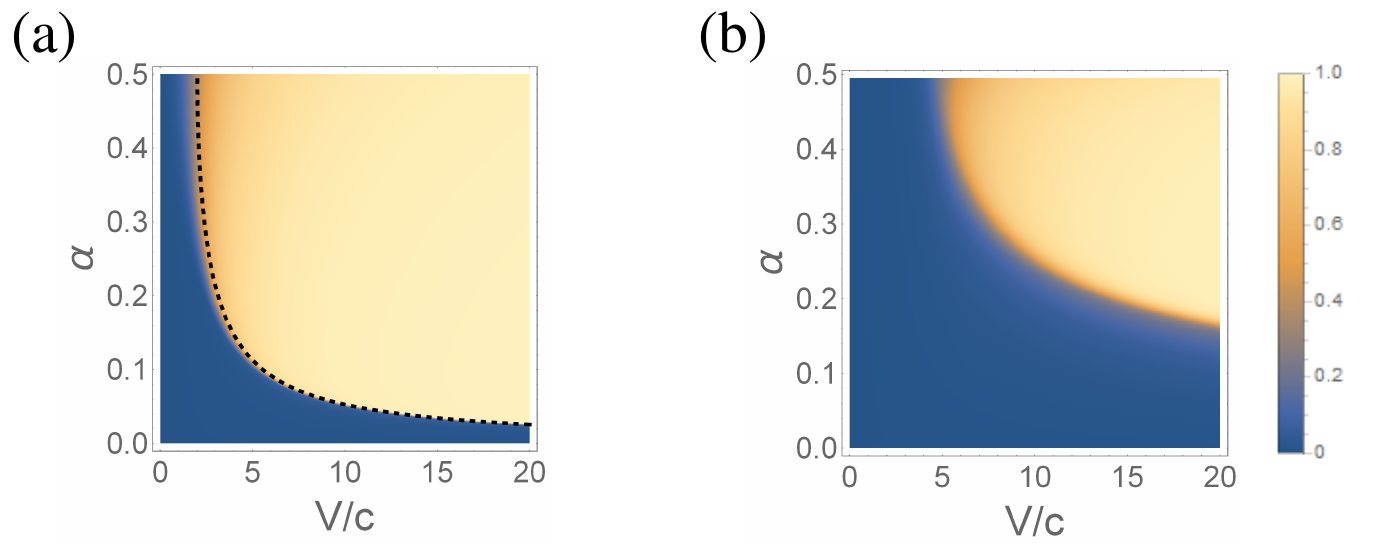}
\caption{\label{fig_locphase} Average eigenstate participation ratio for (a) the 1D and (b) the 2D self-similar structures. The sharp transition indicates a phase transition in which the eigenstates go from being extended to being localized. A $1024$ site chain and $32 \times 32$ site grid were used for the 1D and 2D structures, respectively. In the 1D case, an analytical approximation for the phase boundary can be calculated for small $\alpha$ and yields the boundary shown in panel (a) as a dashed line.}
\end{figure}

To further study the localization properties associated with these potentials, we show in Figure \ref{fig_locphase} the average eigenstate participation ratio as a function of the parameters $V/c$ and $\alpha$. We note the presence of a sharp boundary at which the participation ratio rapidly changes, indicating a phase transition between a phase with extended eigenstates and a phase with localized eigenstates. As shown in the Supplemental Material \cite{suppmat}, an analytical approximation to the phase boundary can be computed in 1D using a renormalization group based calculation and is shown in Figure \ref{fig_locphase}(a) as a dashed line. We contrast these phase transitions with the behavior of disordered systems, which exhibit localization for arbitrarily weak disorder in both 1D and 2D and only exhibit a localization-delocalization transition in 3D. The presence of such phase transitions in lower-dimensional potentials has previously been observed for other aperiodic structures such as the Aubry-Andr\'e model \cite{aubry1980analyticity,Lahini2009}. 

In practice, implementations of an onsite energy configuration will be subject to constraints of finite system size as well as restrictions on how many distinct onsite energies are available within fabrication tolerances. As the self-similar structure described above is constructed via a recursive series of detuning operations\textemdash each of which yields a larger unit cell requiring a larger but finite number of distinct onsite energies\textemdash it provides us with a natural set of periodic structures that can be used to approximate the full structure under a specified set of fabrications constraints. Here we characterize these structures based on the maximal propagation distance at which they are able to maintain the localization properties achieved by the full aperiodic structure. 

\begin{figure}
\centering
\includegraphics[width=1\columnwidth]{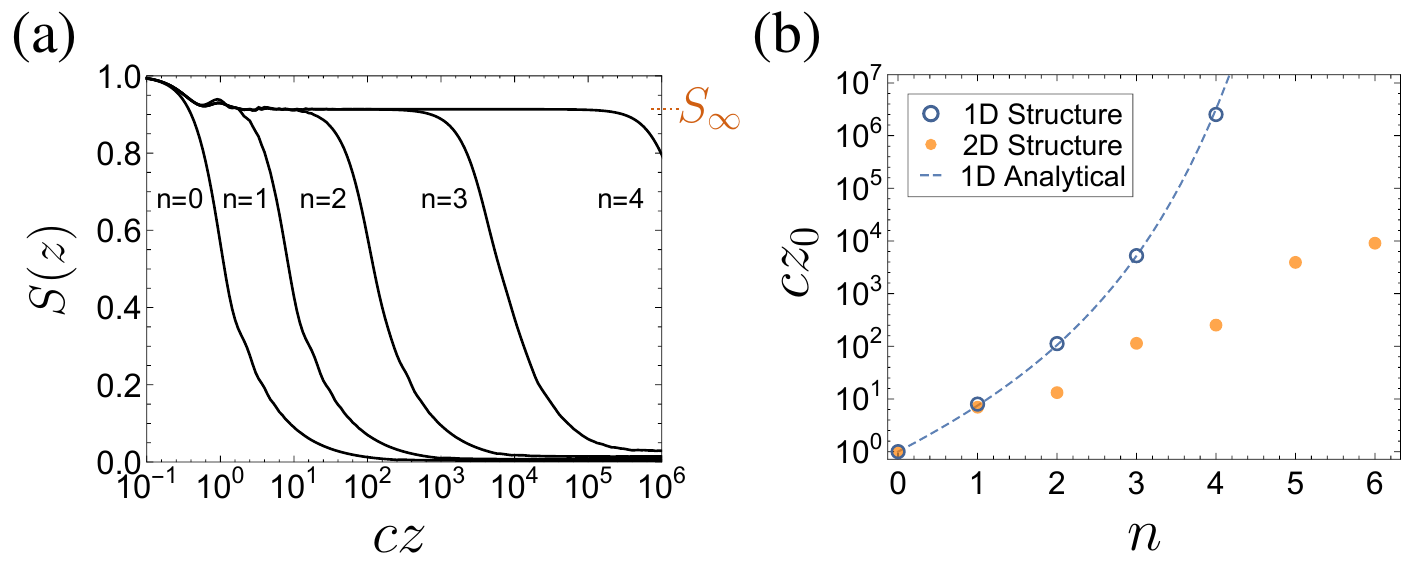}
\caption{\label{fig_ptrunc}Behavior of periodic truncations of the fractal. (a) Finite $z$ objective function plotted for the 1D structure with $\alpha = 0.25$ and $V/c =5$ (parameters which, for the full structure, yield a localized phase) truncated at various detuning levels $n$. The function follows the value $S_\infty$ associated with the full aperiodic structure out to some critical propagation distance $z_0$ which diverges as $n \to \infty$. (b) Dependence of $z_0$ on $n$ for a 1D structure defined by the parameters mentioned above and a 2D structure defined by parameters $\alpha = 0.36$ and $V/c= 10$. In the 1D case, the dashed curve shows the analytical result obtained from the effective coupling $c(N)$ discussed in the text.}
\end{figure}

As the truncated structures are periodic, the eigenstates will be extended and the corresponding value of $S(z\to \infty)$ (i.e., the average participation ratio) vanishes. As a result, a different metric is necessary to study the periodic truncations. We therefore relax our criteria and evaluate $S(z)$ at finite propagation distance. Defining $S_\infty  \equiv \lim_{z\to\infty}S(z)$ evaluated for the full aperiodic structure, a given periodic truncation will produce a function $S(z)$ that approximately follows the value $S_\infty$ out to some critical propagation distance, $z_0$, beyond which it approaches zero. This effect is demonstrated in Figure \ref{fig_ptrunc}(a), where we plot $S(z)$ for the 1D structure truncated at unit cells of size $2^n$ for $n=0,\ldots,4$. The behavior in the 2D case is similar. 

The propagation distance at which the function drops below $S_\infty$ can be understood as the coupling length scale of diffraction associated with the average width of the $2^n$ individual bands of the periodic system. In Figure \ref{fig_ptrunc}(b), we plot $z_0$, as extracted from the average bandwidth, as a function of $n$ for both the 1D and 2D cases. Here we see the power of this structure in that increasing the level $n$ of the truncation results in a drastic increase in the distance over which the structure is capable of suppressing diffraction. We further note that, in the 1D case, we can obtain an analytical approximation for this coupling length. Defining $N \equiv 2^n$ as the number of sites in the unit cell after $n$ detuning operations, the effective coupling $c(N)$ that determines the coupling length takes the form $c(N) = c\,e^{-\gamma(N-1)}/N^\beta$ with $\gamma = \log(2\alpha V_0/c)$, $\beta = \log_2(1/\alpha)$, and $c$ the nearest-neighbor coupling of the physical structure (see Supplemental Material \cite{suppmat}). This result is plotted in Figure \ref{fig_ptrunc}(b) as a dashed line.

\textit{Discussion}.\textemdash Motivated by the ability of disordered and flat-band potentials to reduce diffraction, we have studied the question of how one should structure a potential in order to achieve maximal localization. We conclude that, while disorder is efficient in the sense of being able to induce localization for arbitrarily weak disorder, it is inefficient in the degree of localization it generates. In particular, we have found that, when the range of available onsite energies is large relative to the coupling, maximally efficient localization can be obtained by constructing a highly ordered, self-similar potential in which the onsite energies of neighboring sites are recursively detuned from one another by an amount commensurate to their separation distance.  Like the disordered case, the resulting structure is aperiodic and, as such, circumvents Bloch's theorem, allowing for the existence of localized modes. Unlike disorder, however, it exhibits a localization-delocalization transition in both 1D and 2D. Almost immediately upon passing through the localization transition, the potential induces eigenstate localization that, in comparison to a baseline obtained using simulated annealing, is the maximal possible localization attainable at a fixed potential strength. We also note that, in comparison to approaches based on engineered flat bands, our structure is advantageous in that it avoids the bending losses associated with a Floquet drive and does not possess the additional dispersive bands associated with static flat-band systems like the Lieb lattice.

A useful feature of these potentials arises from their construction via a series of periodic structures with increasingly larger unit cells. These periodic structures are capable of suppressing diffraction comparably to the full aperiodic structures out to a propagation distance that diverges rapidly with the unit cell size\textemdash a feature that could find use for crosstalk reduction in spatially multiplexed fibers used in long-haul data transmission. The limiting aperiodic structure both provides a direct recipe for designing propagation constant configurations that suppress diffraction at a given required propagation distance, as well as clarifies the existence of a localization-delocalization transition that affects the ability of the periodic truncations to suppress diffraction at long distances.

\begin{acknowledgments}
M.C.R. acknowledges the National Science Foundation under award number ECCS-1509546, the Charles E. Kaufman Foundation (a supporting organization of the Pittsburgh Foundation) under grant number KA2017-91788, the Packard foundation under fellowship number 2017-66821, and the Alfred P. Sloan Foundation under fellowship number FG-2016-6418.
\end{acknowledgments}

\bibliography{imagingrefs}

\begin{thebibliography}{21}%
\makeatletter
\providecommand \@ifxundefined [1]{%
 \@ifx{#1\undefined}
}%
\providecommand \@ifnum [1]{%
 \ifnum #1\expandafter \@firstoftwo
 \else \expandafter \@secondoftwo
 \fi
}%
\providecommand \@ifx [1]{%
 \ifx #1\expandafter \@firstoftwo
 \else \expandafter \@secondoftwo
 \fi
}%
\providecommand \natexlab [1]{#1}%
\providecommand \enquote  [1]{``#1''}%
\providecommand \bibnamefont  [1]{#1}%
\providecommand \bibfnamefont [1]{#1}%
\providecommand \citenamefont [1]{#1}%
\providecommand \href@noop [0]{\@secondoftwo}%
\providecommand \href [0]{\begingroup \@sanitize@url \@href}%
\providecommand \@href[1]{\@@startlink{#1}\@@href}%
\providecommand \@@href[1]{\endgroup#1\@@endlink}%
\providecommand \@sanitize@url [0]{\catcode `\\12\catcode `\$12\catcode
  `\&12\catcode `\#12\catcode `\^12\catcode `\_12\catcode `\%12\relax}%
\providecommand \@@startlink[1]{}%
\providecommand \@@endlink[0]{}%
\providecommand \url  [0]{\begingroup\@sanitize@url \@url }%
\providecommand \@url [1]{\endgroup\@href {#1}{\urlprefix }}%
\providecommand \urlprefix  [0]{URL }%
\providecommand \Eprint [0]{\href }%
\providecommand \doibase [0]{http://dx.doi.org/}%
\providecommand \selectlanguage [0]{\@gobble}%
\providecommand \bibinfo  [0]{\@secondoftwo}%
\providecommand \bibfield  [0]{\@secondoftwo}%
\providecommand \translation [1]{[#1]}%
\providecommand \BibitemOpen [0]{}%
\providecommand \bibitemStop [0]{}%
\providecommand \bibitemNoStop [0]{.\EOS\space}%
\providecommand \EOS [0]{\spacefactor3000\relax}%
\providecommand \BibitemShut  [1]{\csname bibitem#1\endcsname}%
\let\auto@bib@innerbib\@empty
\bibitem [{\citenamefont {Hopkins}\ and\ \citenamefont
  {Kapany}(1954)}]{hopkins1954}%
  \BibitemOpen
  \bibfield  {author} {\bibinfo {author} {\bibfnamefont {H.~H.}\ \bibnamefont
  {Hopkins}}\ and\ \bibinfo {author} {\bibfnamefont {N.~S.}\ \bibnamefont
  {Kapany}},\ }\href@noop {} {\bibfield  {journal} {\bibinfo  {journal}
  {Nature}\ }\textbf {\bibinfo {volume} {173}},\ \bibinfo {pages} {39}
  (\bibinfo {year} {1954})}\BibitemShut {NoStop}%
\bibitem [{\citenamefont {Richardson}\ \emph {et~al.}(2013)\citenamefont
  {Richardson}, \citenamefont {Fini},\ and\ \citenamefont
  {Nelson}}]{Richardson2013}%
  \BibitemOpen
  \bibfield  {author} {\bibinfo {author} {\bibfnamefont {D.~J.}\ \bibnamefont
  {Richardson}}, \bibinfo {author} {\bibfnamefont {J.~M.}\ \bibnamefont
  {Fini}}, \ and\ \bibinfo {author} {\bibfnamefont {L.~E.}\ \bibnamefont
  {Nelson}},\ }\href@noop {} {\bibfield  {journal} {\bibinfo  {journal} {Nature
  Photonics}\ }\textbf {\bibinfo {volume} {7}},\ \bibinfo {pages} {354}
  (\bibinfo {year} {2013})}\BibitemShut {NoStop}%
\bibitem [{\citenamefont {Winzer}(2014)}]{Winzer2014}%
  \BibitemOpen
  \bibfield  {author} {\bibinfo {author} {\bibfnamefont {P.~J.}\ \bibnamefont
  {Winzer}},\ }\href@noop {} {\bibfield  {journal} {\bibinfo  {journal} {Bell
  Labs Technical Journal}\ }\textbf {\bibinfo {volume} {19}},\ \bibinfo {pages}
  {22} (\bibinfo {year} {2014})}\BibitemShut {NoStop}%
\bibitem [{\citenamefont {Eisenberg}\ \emph {et~al.}(2000)\citenamefont
  {Eisenberg}, \citenamefont {Silberberg}, \citenamefont {Morandotti},\ and\
  \citenamefont {Aitchison}}]{Eisenberg2000}%
  \BibitemOpen
  \bibfield  {author} {\bibinfo {author} {\bibfnamefont {H.~S.}\ \bibnamefont
  {Eisenberg}}, \bibinfo {author} {\bibfnamefont {Y.}~\bibnamefont
  {Silberberg}}, \bibinfo {author} {\bibfnamefont {R.}~\bibnamefont
  {Morandotti}}, \ and\ \bibinfo {author} {\bibfnamefont {J.~S.}\ \bibnamefont
  {Aitchison}},\ }\href {\doibase 10.1103/PhysRevLett.85.1863} {\bibfield
  {journal} {\bibinfo  {journal} {Phys. Rev. Lett.}\ }\textbf {\bibinfo
  {volume} {85}},\ \bibinfo {pages} {1863} (\bibinfo {year}
  {2000})}\BibitemShut {NoStop}%
\bibitem [{\citenamefont {Longhi}\ \emph {et~al.}(2006)\citenamefont {Longhi},
  \citenamefont {Marangoni}, \citenamefont {Lobino}, \citenamefont {Ramponi},
  \citenamefont {Laporta}, \citenamefont {Cianci},\ and\ \citenamefont
  {Foglietti}}]{Longhi2006}%
  \BibitemOpen
  \bibfield  {author} {\bibinfo {author} {\bibfnamefont {S.}~\bibnamefont
  {Longhi}}, \bibinfo {author} {\bibfnamefont {M.}~\bibnamefont {Marangoni}},
  \bibinfo {author} {\bibfnamefont {M.}~\bibnamefont {Lobino}}, \bibinfo
  {author} {\bibfnamefont {R.}~\bibnamefont {Ramponi}}, \bibinfo {author}
  {\bibfnamefont {P.}~\bibnamefont {Laporta}}, \bibinfo {author} {\bibfnamefont
  {E.}~\bibnamefont {Cianci}}, \ and\ \bibinfo {author} {\bibfnamefont
  {V.}~\bibnamefont {Foglietti}},\ }\href {\doibase
  10.1103/PhysRevLett.96.243901} {\bibfield  {journal} {\bibinfo  {journal}
  {Phys. Rev. Lett.}\ }\textbf {\bibinfo {volume} {96}},\ \bibinfo {pages}
  {243901} (\bibinfo {year} {2006})}\BibitemShut {NoStop}%
\bibitem [{\citenamefont {Szameit}\ \emph {et~al.}(2010)\citenamefont
  {Szameit}, \citenamefont {Garanovich}, \citenamefont {Heinrich},
  \citenamefont {Sukhorukov}, \citenamefont {Dreisow}, \citenamefont {Pertsch},
  \citenamefont {Nolte}, \citenamefont {T\"unnermann}, \citenamefont {Longhi},\
  and\ \citenamefont {Kivshar}}]{Szameit2010}%
  \BibitemOpen
  \bibfield  {author} {\bibinfo {author} {\bibfnamefont {A.}~\bibnamefont
  {Szameit}}, \bibinfo {author} {\bibfnamefont {I.~L.}\ \bibnamefont
  {Garanovich}}, \bibinfo {author} {\bibfnamefont {M.}~\bibnamefont
  {Heinrich}}, \bibinfo {author} {\bibfnamefont {A.~A.}\ \bibnamefont
  {Sukhorukov}}, \bibinfo {author} {\bibfnamefont {F.}~\bibnamefont {Dreisow}},
  \bibinfo {author} {\bibfnamefont {T.}~\bibnamefont {Pertsch}}, \bibinfo
  {author} {\bibfnamefont {S.}~\bibnamefont {Nolte}}, \bibinfo {author}
  {\bibfnamefont {A.}~\bibnamefont {T\"unnermann}}, \bibinfo {author}
  {\bibfnamefont {S.}~\bibnamefont {Longhi}}, \ and\ \bibinfo {author}
  {\bibfnamefont {Y.~S.}\ \bibnamefont {Kivshar}},\ }\href {\doibase
  10.1103/PhysRevLett.104.223903} {\bibfield  {journal} {\bibinfo  {journal}
  {Phys. Rev. Lett.}\ }\textbf {\bibinfo {volume} {104}},\ \bibinfo {pages}
  {223903} (\bibinfo {year} {2010})}\BibitemShut {NoStop}%
\bibitem [{\citenamefont {Garanovich}\ \emph {et~al.}(2012)\citenamefont
  {Garanovich}, \citenamefont {Longhi}, \citenamefont {Sukhorukov},\ and\
  \citenamefont {Kivshar}}]{Garanovich2012}%
  \BibitemOpen
  \bibfield  {author} {\bibinfo {author} {\bibfnamefont {I.~L.}\ \bibnamefont
  {Garanovich}}, \bibinfo {author} {\bibfnamefont {S.}~\bibnamefont {Longhi}},
  \bibinfo {author} {\bibfnamefont {A.~A.}\ \bibnamefont {Sukhorukov}}, \ and\
  \bibinfo {author} {\bibfnamefont {Y.~S.}\ \bibnamefont {Kivshar}},\ }\href
  {\doibase https://doi.org/10.1016/j.physrep.2012.03.005} {\bibfield
  {journal} {\bibinfo  {journal} {Physics Reports}\ }\textbf {\bibinfo {volume}
  {518}},\ \bibinfo {pages} {1 } (\bibinfo {year} {2012})}\BibitemShut
  {NoStop}%
\bibitem [{\citenamefont {Crespi}\ \emph {et~al.}(2013)\citenamefont {Crespi},
  \citenamefont {Corrielli}, \citenamefont {Valle}, \citenamefont {Osellame},\
  and\ \citenamefont {Longhi}}]{Crespi2013}%
  \BibitemOpen
  \bibfield  {author} {\bibinfo {author} {\bibfnamefont {A.}~\bibnamefont
  {Crespi}}, \bibinfo {author} {\bibfnamefont {G.}~\bibnamefont {Corrielli}},
  \bibinfo {author} {\bibfnamefont {G.~D.}\ \bibnamefont {Valle}}, \bibinfo
  {author} {\bibfnamefont {R.}~\bibnamefont {Osellame}}, \ and\ \bibinfo
  {author} {\bibfnamefont {S.}~\bibnamefont {Longhi}},\ }\href@noop {}
  {\bibfield  {journal} {\bibinfo  {journal} {New Journal of Physics}\ }\textbf
  {\bibinfo {volume} {15}},\ \bibinfo {pages} {013012} (\bibinfo {year}
  {2013})}\BibitemShut {NoStop}%
\bibitem [{\citenamefont {Vicencio}\ and\ \citenamefont
  {Mejía-Cortés}(2014)}]{Vicencio2014}%
  \BibitemOpen
  \bibfield  {author} {\bibinfo {author} {\bibfnamefont {R.~A.}\ \bibnamefont
  {Vicencio}}\ and\ \bibinfo {author} {\bibfnamefont {C.}~\bibnamefont
  {Mejía-Cortés}},\ }\href@noop {} {\bibfield  {journal} {\bibinfo  {journal}
  {Journal of Optics}\ }\textbf {\bibinfo {volume} {16}},\ \bibinfo {pages}
  {015706} (\bibinfo {year} {2014})}\BibitemShut {NoStop}%
\bibitem [{\citenamefont {Guzmán-Silva}\ \emph {et~al.}(2014)\citenamefont
  {Guzmán-Silva}, \citenamefont {Mejía-Cortés}, \citenamefont {Bandres},
  \citenamefont {Rechtsman}, \citenamefont {Weimann}, \citenamefont {Nolte},
  \citenamefont {Segev}, \citenamefont {Szameit},\ and\ \citenamefont
  {Vicencio}}]{Silva2014}%
  \BibitemOpen
  \bibfield  {author} {\bibinfo {author} {\bibfnamefont {D.}~\bibnamefont
  {Guzmán-Silva}}, \bibinfo {author} {\bibfnamefont {C.}~\bibnamefont
  {Mejía-Cortés}}, \bibinfo {author} {\bibfnamefont {M.~A.}\ \bibnamefont
  {Bandres}}, \bibinfo {author} {\bibfnamefont {M.~C.}\ \bibnamefont
  {Rechtsman}}, \bibinfo {author} {\bibfnamefont {S.}~\bibnamefont {Weimann}},
  \bibinfo {author} {\bibfnamefont {S.}~\bibnamefont {Nolte}}, \bibinfo
  {author} {\bibfnamefont {M.}~\bibnamefont {Segev}}, \bibinfo {author}
  {\bibfnamefont {A.}~\bibnamefont {Szameit}}, \ and\ \bibinfo {author}
  {\bibfnamefont {R.~A.}\ \bibnamefont {Vicencio}},\ }\href@noop {} {\bibfield
  {journal} {\bibinfo  {journal} {New Journal of Physics}\ }\textbf {\bibinfo
  {volume} {16}},\ \bibinfo {pages} {063061} (\bibinfo {year}
  {2014})}\BibitemShut {NoStop}%
\bibitem [{\citenamefont {Mukherjee}\ \emph {et~al.}(2015)\citenamefont
  {Mukherjee}, \citenamefont {Spracklen}, \citenamefont {Choudhury},
  \citenamefont {Goldman}, \citenamefont {\"Ohberg}, \citenamefont
  {Andersson},\ and\ \citenamefont {Thomson}}]{Mukherjee2015}%
  \BibitemOpen
  \bibfield  {author} {\bibinfo {author} {\bibfnamefont {S.}~\bibnamefont
  {Mukherjee}}, \bibinfo {author} {\bibfnamefont {A.}~\bibnamefont
  {Spracklen}}, \bibinfo {author} {\bibfnamefont {D.}~\bibnamefont
  {Choudhury}}, \bibinfo {author} {\bibfnamefont {N.}~\bibnamefont {Goldman}},
  \bibinfo {author} {\bibfnamefont {P.}~\bibnamefont {\"Ohberg}}, \bibinfo
  {author} {\bibfnamefont {E.}~\bibnamefont {Andersson}}, \ and\ \bibinfo
  {author} {\bibfnamefont {R.~R.}\ \bibnamefont {Thomson}},\ }\href {\doibase
  10.1103/PhysRevLett.114.245504} {\bibfield  {journal} {\bibinfo  {journal}
  {Phys. Rev. Lett.}\ }\textbf {\bibinfo {volume} {114}},\ \bibinfo {pages}
  {245504} (\bibinfo {year} {2015})}\BibitemShut {NoStop}%
\bibitem [{\citenamefont {Koshiba}\ \emph {et~al.}(2009)\citenamefont
  {Koshiba}, \citenamefont {Saitoh},\ and\ \citenamefont
  {Kokubun}}]{Koshiba2009}%
  \BibitemOpen
  \bibfield  {author} {\bibinfo {author} {\bibfnamefont {M.}~\bibnamefont
  {Koshiba}}, \bibinfo {author} {\bibfnamefont {K.}~\bibnamefont {Saitoh}}, \
  and\ \bibinfo {author} {\bibfnamefont {Y.}~\bibnamefont {Kokubun}},\
  }\href@noop {} {\bibfield  {journal} {\bibinfo  {journal} {IEICE Electronics
  Express}\ }\textbf {\bibinfo {volume} {6}},\ \bibinfo {pages} {98} (\bibinfo
  {year} {2009})}\BibitemShut {NoStop}%
\bibitem [{\citenamefont {Anderson}(1958)}]{AndLoc1}%
  \BibitemOpen
  \bibfield  {author} {\bibinfo {author} {\bibfnamefont {P.~W.}\ \bibnamefont
  {Anderson}},\ }\href@noop {} {\bibfield  {journal} {\bibinfo  {journal}
  {Physical review}\ }\textbf {\bibinfo {volume} {109}},\ \bibinfo {pages}
  {1492} (\bibinfo {year} {1958})}\BibitemShut {NoStop}%
\bibitem [{\citenamefont {Schwartz}\ \emph {et~al.}(2007)\citenamefont
  {Schwartz}, \citenamefont {Bartal}, \citenamefont {Fishman},\ and\
  \citenamefont {Segev}}]{AndLoc5}%
  \BibitemOpen
  \bibfield  {author} {\bibinfo {author} {\bibfnamefont {T.}~\bibnamefont
  {Schwartz}}, \bibinfo {author} {\bibfnamefont {G.}~\bibnamefont {Bartal}},
  \bibinfo {author} {\bibfnamefont {S.}~\bibnamefont {Fishman}}, \ and\
  \bibinfo {author} {\bibfnamefont {M.}~\bibnamefont {Segev}},\ }\href@noop {}
  {\bibfield  {journal} {\bibinfo  {journal} {Nature}\ }\textbf {\bibinfo
  {volume} {446}},\ \bibinfo {pages} {52} (\bibinfo {year} {2007})}\BibitemShut
  {NoStop}%
\bibitem [{\citenamefont {Karbasi}\ \emph
  {et~al.}(2012{\natexlab{a}})\citenamefont {Karbasi}, \citenamefont {Mirr},
  \citenamefont {Yarandi}, \citenamefont {Frazier}, \citenamefont {Koch},\ and\
  \citenamefont {Mafi}}]{Mafi1}%
  \BibitemOpen
  \bibfield  {author} {\bibinfo {author} {\bibfnamefont {S.}~\bibnamefont
  {Karbasi}}, \bibinfo {author} {\bibfnamefont {C.~R.}\ \bibnamefont {Mirr}},
  \bibinfo {author} {\bibfnamefont {P.~G.}\ \bibnamefont {Yarandi}}, \bibinfo
  {author} {\bibfnamefont {R.~J.}\ \bibnamefont {Frazier}}, \bibinfo {author}
  {\bibfnamefont {K.~W.}\ \bibnamefont {Koch}}, \ and\ \bibinfo {author}
  {\bibfnamefont {A.}~\bibnamefont {Mafi}},\ }\href@noop {} {\bibfield
  {journal} {\bibinfo  {journal} {Optics letters}\ }\textbf {\bibinfo {volume}
  {37}},\ \bibinfo {pages} {2304} (\bibinfo {year}
  {2012}{\natexlab{a}})}\BibitemShut {NoStop}%
\bibitem [{\citenamefont {Karbasi}\ \emph {et~al.}(2014)\citenamefont
  {Karbasi}, \citenamefont {Frazier}, \citenamefont {Koch}, \citenamefont
  {Hawkins}, \citenamefont {Ballato},\ and\ \citenamefont {Mafi}}]{Mafi2}%
  \BibitemOpen
  \bibfield  {author} {\bibinfo {author} {\bibfnamefont {S.}~\bibnamefont
  {Karbasi}}, \bibinfo {author} {\bibfnamefont {R.~J.}\ \bibnamefont
  {Frazier}}, \bibinfo {author} {\bibfnamefont {K.~W.}\ \bibnamefont {Koch}},
  \bibinfo {author} {\bibfnamefont {T.}~\bibnamefont {Hawkins}}, \bibinfo
  {author} {\bibfnamefont {J.}~\bibnamefont {Ballato}}, \ and\ \bibinfo
  {author} {\bibfnamefont {A.}~\bibnamefont {Mafi}},\ }\href@noop {} {\bibfield
   {journal} {\bibinfo  {journal} {Nature Communications}\ }\textbf {\bibinfo
  {volume} {5}},\ \bibinfo {pages} {3362} (\bibinfo {year} {2014})}\BibitemShut
  {NoStop}%
\bibitem [{\citenamefont {Karbasi}\ \emph
  {et~al.}(2012{\natexlab{b}})\citenamefont {Karbasi}, \citenamefont {Hawkins},
  \citenamefont {Ballato}, \citenamefont {Koch},\ and\ \citenamefont
  {Mafi}}]{Mafi3}%
  \BibitemOpen
  \bibfield  {author} {\bibinfo {author} {\bibfnamefont {S.}~\bibnamefont
  {Karbasi}}, \bibinfo {author} {\bibfnamefont {T.}~\bibnamefont {Hawkins}},
  \bibinfo {author} {\bibfnamefont {J.}~\bibnamefont {Ballato}}, \bibinfo
  {author} {\bibfnamefont {K.~W.}\ \bibnamefont {Koch}}, \ and\ \bibinfo
  {author} {\bibfnamefont {A.}~\bibnamefont {Mafi}},\ }\href@noop {} {\bibfield
   {journal} {\bibinfo  {journal} {Optical Materials Express}\ }\textbf
  {\bibinfo {volume} {2}},\ \bibinfo {pages} {1496} (\bibinfo {year}
  {2012}{\natexlab{b}})}\BibitemShut {NoStop}%
\bibitem [{sup()}]{suppmat}%
  \BibitemOpen
  \href@noop {} {}\bibinfo {note} {See Supplemental Material for additional
  theoretical derivations; see ancillary files for animations showing
  simulations of image propagation through the fractal and disordered
  potentials.}\BibitemShut {Stop}%
\bibitem [{\citenamefont {Goodman}(2017)}]{goodman2017}%
  \BibitemOpen
  \bibfield  {author} {\bibinfo {author} {\bibfnamefont {J.}~\bibnamefont
  {Goodman}},\ }\href@noop {} {\emph {\bibinfo {title} {Introduction to Fourier
  Optics}}}\ (\bibinfo  {publisher} {W. H. Freeman},\ \bibinfo {year}
  {2017})\BibitemShut {NoStop}%
\bibitem [{\citenamefont {Aubry}\ and\ \citenamefont
  {Andr{\'e}}(1980)}]{aubry1980analyticity}%
  \BibitemOpen
  \bibfield  {author} {\bibinfo {author} {\bibfnamefont {S.}~\bibnamefont
  {Aubry}}\ and\ \bibinfo {author} {\bibfnamefont {G.}~\bibnamefont
  {Andr{\'e}}},\ }\href@noop {} {\bibfield  {journal} {\bibinfo  {journal}
  {Ann. Isr. Phys. Soc.}\ }\textbf {\bibinfo {volume} {3}},\ \bibinfo {pages}
  {18} (\bibinfo {year} {1980})}\BibitemShut {NoStop}%
\bibitem [{\citenamefont {Lahini}\ \emph {et~al.}(2009)\citenamefont {Lahini},
  \citenamefont {Pugatch}, \citenamefont {Pozzi}, \citenamefont {Sorel},
  \citenamefont {Morandotti}, \citenamefont {Davidson},\ and\ \citenamefont
  {Silberberg}}]{Lahini2009}%
  \BibitemOpen
  \bibfield  {author} {\bibinfo {author} {\bibfnamefont {Y.}~\bibnamefont
  {Lahini}}, \bibinfo {author} {\bibfnamefont {R.}~\bibnamefont {Pugatch}},
  \bibinfo {author} {\bibfnamefont {F.}~\bibnamefont {Pozzi}}, \bibinfo
  {author} {\bibfnamefont {M.}~\bibnamefont {Sorel}}, \bibinfo {author}
  {\bibfnamefont {R.}~\bibnamefont {Morandotti}}, \bibinfo {author}
  {\bibfnamefont {N.}~\bibnamefont {Davidson}}, \ and\ \bibinfo {author}
  {\bibfnamefont {Y.}~\bibnamefont {Silberberg}},\ }\href {\doibase
  10.1103/PhysRevLett.103.013901} {\bibfield  {journal} {\bibinfo  {journal}
  {Phys. Rev. Lett.}\ }\textbf {\bibinfo {volume} {103}},\ \bibinfo {pages}
  {013901} (\bibinfo {year} {2009})}\BibitemShut {NoStop}%
\end{thebibliography}%

\newpage
\onecolumngrid
\begin{center}
\large
\textbf{Supplemental material: Fractal waveguide arrays induce maximal localization}
\end{center}

\begin{figure}
\centering
\includegraphics[width=0.9\columnwidth]{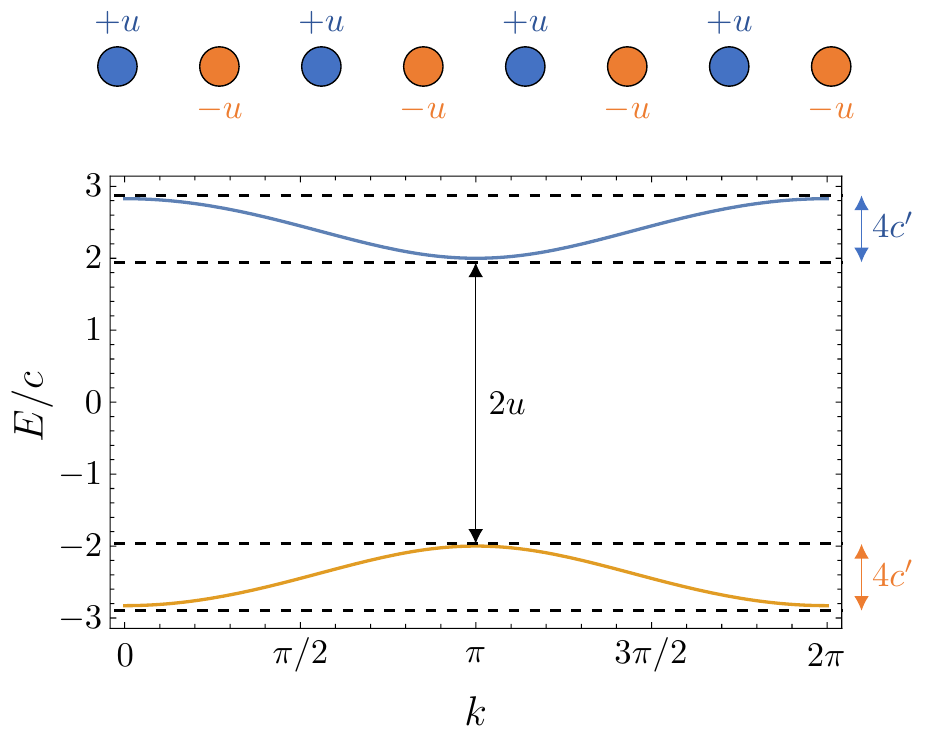}
\caption{\label{fig_lineardetuned} Illustration of a detuned linear array and its band structure. In the strongly detuned regime, the upper and lower bands become highly localized on the $+u$ and $-u$ sublattices, respectively. In this limit, we may reinterpret the two bands as forming two independent arrays governed by a modified interwaveguide coupling $c'$ that is determined by the width of the individual bands.}
\end{figure}

\twocolumngrid
\textit{Analytical approximation to phase boundary}.\textemdash Here we compute an analytical approximation for the localization-delocalization phase boundary for the 1D self-similar structure discussed in the text. We begin by considering an equally spaced 1D array in which adjacent sites are detuned by $\pm u$ (Figure \ref{fig_lineardetuned}). Assuming a nearest-neighbor coupling of $c$, the eigenstates form two bands with energies given by
\begin{align}
E(k) &= \pm \sqrt{u^2 + 4 c^2\cos^2(k/2)}\\
&=\pm(u^2+2c^2)^{1/2} \sqrt{1+\frac{2c^2}{u^2+2c^2}\cos k}
\end{align}
which can be expanded as
\begin{equation}
E(k) = \pm\left[(u^2+2c^2)^{1/2} + \frac{c^2\cos k}{(u^2+2c^2)^{1/2}} + \ldots\right]
\end{equation}
We now consider the regime where $c/u \ll 1$ and neglect higher-order terms in the expansion. Noting that, in this approximation, the individual bands obey a dispersion proportional to $\cos(k)$ (up to an offset in energy), we may reinterpret the two bands as forming two distinct linear arrays with a modified nearest-neighbor coupling determined by the bandwidth: 
\begin{align}
c' &= \frac{c^2}{2(u^2+2c^2)^{1/2}}\\
&= c^2/(2 u) + O(c^3/u^3).
\end{align}

We now make an additional approximation in which we note that, in the regime $c/u \ll 1$, the eigenstates of the upper and lower bands are highly localized on the $+u$ and $-u$ sublattices, respectively. Since the construction of the potential described in the text proceeds by independently reapplying the detuning procedure on these two sublattices (with smaller detuning energies $\pm\alpha u$), we may iterate the procedure discussed above to generate a flow of the effective interwaveguide coupling. In particular, for the next iteration, we repeat the above calculation using a new coupling $c' = c^2/(2u)$ and a new detuning $u' = \alpha u$. This will again generate a modified coupling $c'' = c'^2/(2u') = c^4/(8\alpha u^3)$. More generally, we define $c_k$ and $u_k$ as the effective coupling and energy detuning at recursion level $k$. In particular, for the structure defined in the text $u_k = \alpha^{k} V_0$. 

Note that, in the approximation introduced above, the eigenstates will localize when the ratio $r_k \equiv c_k/u_k$ goes to zero for large $k$, as this indicates that the asymptotic energy detuning becomes infinitely large relative to the corresponding effective coupling. From the procedure introduced above, we have $c_{k+1} = c^2_k/(2u_k)$ and hence
\begin{equation}
r_{k+1} = \left(\frac{u_k}{2u_{k+1}}\right) r_k^2.
\end{equation}
This recursion relation can be solved exactly, yielding 
\begin{equation}
\log r_k = 2^k \left\{ \sum_{m=1}^k \frac{1}{2^m} \log\left(\frac{u_{m-1}}{2\,u_m}\right)+ \log r_0\right\}.
\end{equation}
Using the detuning function $u_k = \alpha^k V_0$ and denoting $c_0 = c$, this simplifies to 
\begin{equation}\label{eqn_ratio}
\log r_k = 2^k \log\left(\frac{c}{2\alpha\,  V_0}\right)- \log\left(\frac{1}{2\alpha}\right)
\end{equation}
so that $\lim_{k\to\infty}r_k\ = 0$ when
\begin{equation}\label{eqn_phase_bound}
\frac{V_0}{c} > \frac{1}{2\alpha},
\end{equation}
indicating eigenstate localization. In the opposite regime, $V_0/c < 1/(2\alpha)$, the ratios formally diverge, $\lim_{k\to\infty}r_k =  \infty$, indicating that the asymptotic onsite energies become negligible relative to the corresponding couplings, suggesting the presence of extended eigenstates. We note, however, that in this latter regime, the expansion used above with $c/u \ll 1$ begins to break down after some finite number of iterations $k$. Hence, this regime must ultimately be explored numerically. Finally, to keep the derivation consistent with the assumption $c/u \ll 1$, we require  $\alpha$ appearing in Equation \ref{eqn_phase_bound} to be small. In this regime, the bound is in good agreement with the numerical results, as shown in the localization-delocalization phase diagram of the main text (which shows the boundary curve defined by Equation (\ref{eqn_phase_bound}) after using the relation $V = V_0/(1-\alpha)$). 

We note that Equation \ref{eqn_ratio} also contains information about the critical propagation distances, $z_0$, which, as discussed in the text, determine the maximal distance at which the periodic truncations of the structure are capable of suppressing diffraction. In particular, a periodic structure truncated after $k$ detuning levels has an associated coupling length $z_0 = 1/c_k$ and from Equation \ref{eqn_ratio} we have
\begin{equation}
\log \left(\frac{c_k}{c}\right) = (2^k -1) \log\left(\frac{c}{2\alpha V_0}\right) - k \log\left(\frac{1}{\alpha}\right).
\end{equation} 
Noting that after $k$ detuning operations, the unit cell contains $N=2^k$ sites, we rewrite the coupling as a function of system size via $c_k \to c(N)$ with $N=2^k$ yielding
\begin{equation}
c(N) = c\,e^{-\gamma(N-1)}/N^\beta
\end{equation}
with $\gamma = \log(2\alpha V_0/c)$ and $\beta = \log_2(1/\alpha)$. Note that this result is only applicable to the localized phase where we have $\gamma >0$. The formal divergence of $c(N)$ in the delocalized phase is unphysical and signals the breakdown of the expansion used above. In particular, in the delocalized phase, the $c(N)$ decay slower than the $u_k$, so that eventually the condition $c/u \ll 1$ fails to be satisfied.

\textit{Cantor set associated with the onsite energies}.\textemdash The onsite energies that appear in the construction of the fractal potential lie in a Cantor set determined by the parameter $\alpha$. Here we describe this set for the general case, alluded to in the main text, where the potential is constructed on an $n$-partite lattice that satisfies the property that the $n$ independent sublattices form rescaled copies of the original lattice. The construction proceeds analogously to the bi-partite case. In particular, we begin by choosing an energy scale, $V_0$, and a parameter $\alpha \in [0,\frac{1}{n}]$. We then divide the interval $[-V_0,+V_0]$  into $n$ equally spaced onsite energies, $\{v_1,...,v_n\}$, and assign these onsite energies to the $n$ sublattices so as to detune neighboring sites. This constitutes the $n=1$ level structure. To generate the $n=2$ level structure, we repeat this procedure independently on each sublattice, this time detuning over a smaller interval $[ v_i-\alpha V_0 , v_i+ \alpha V_0 ]$ centered on the $v_i$ associated with the respective sublattice. This process is repeated infinitely many times to yield the final onsite energy configuration.

 \begin{figure}
\centering
\includegraphics[width=1\columnwidth]{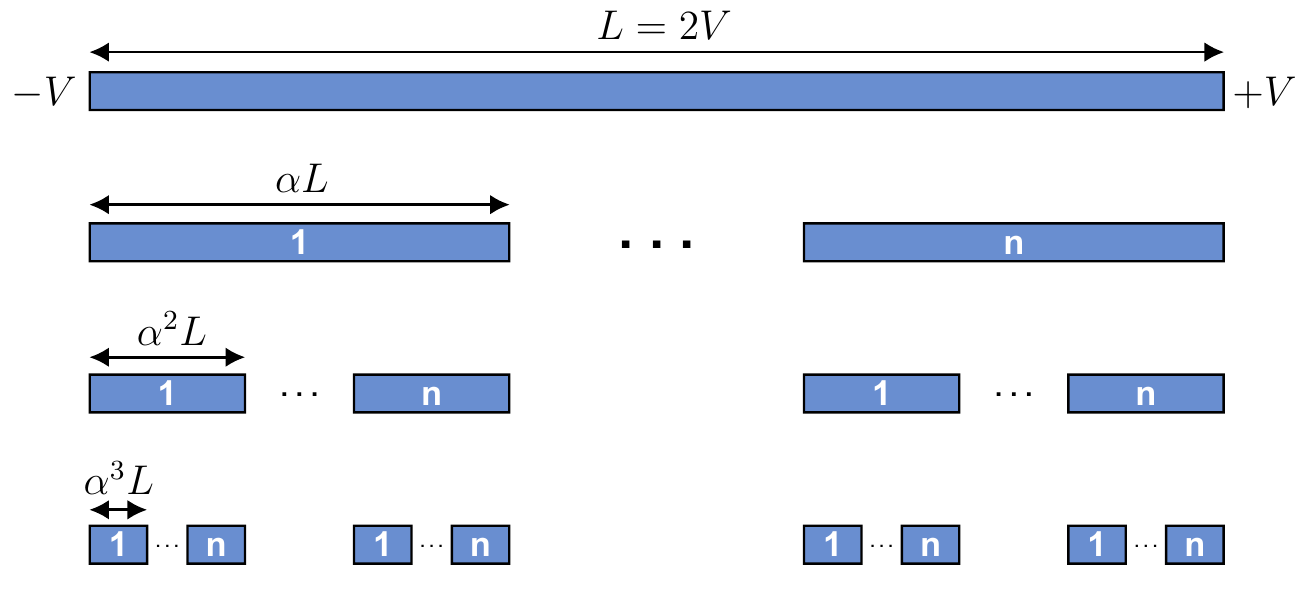}
\caption{\label{fig_cantorset}  Cantor set associated with the onsite energies of the self-similar structure introduced in the text. For the general $n$-partite case, the energy interval $[-V,+V]$ of length $L = 2V$ is divided into $n$ smaller, equally spaced subintervals, each of length $\alpha L$. The gaps between these subintervals represent excluded ranges of onsite energies that are not used in the structure. Iterating this process of subdivision yields a Cantor set within which the onsite energies reside.}
\end{figure}

Note that iterating the above process yields\textemdash for the onsite energies centered on $v_i$\textemdash a bounded set contained in the interval
\begin{equation}
[v_i - \sum_{k=1}^\infty \alpha^k V_0,v_i + \sum_{k=1}^\infty \alpha^k V_0] = [v_i - \alpha V, v_i + \alpha V]
\end{equation}  
where $V = V_0/(1-\alpha)$. Note that this interval has length $\alpha L$ with $L = 2V$. In particular, the procedure has divided the full interval $[-V,+V]$ of length $L=2V$ into $n$ equally spaced, closed subintervals of length $\alpha L$ such that all onsite energies of the final structure lie within these subintervals. Note that the subintervals generally do not cover the full interval $[-V,+V]$ since they have a combined length of $n(\alpha L)$ which, for $\alpha \in[0,\frac{1}{n}]$, is less than or equal to $L$. The spaces between the subintervals form regions of excluded energies that are not assigned to any site in the structure. This is illustrated in Figure \ref{fig_cantorset}. Iterating this argument over higher levels of detuning yields a Cantor set within which the onsite energies of the final structure reside. Note that, as the collection of onsite energies is countable (since the underlying lattice is countable), the actual onsite energies only form a proper subset of this Cantor set. Finally, we note that this construction makes clear why the general $n$-partite structure is properly parameterized by $\alpha \in \left[0,\frac{1}{n}\right]$: the boundary case, $\alpha = \frac{1}{n}$, yields a subdivision of the interval $[-V,V]$ into subintervals that fully span the interval (i.e., which leave no excluded regions).

\textit{Expressions for $S(z)$ at finite and infinite $z$}.\textemdash We now provide a useful formula for the objective function, $S(z)$, at finite $z$ and derive the relation of its  infinite distance limit to the average eigenstate participation ratio. Consider a Hamiltonian $H$ that is diagonalized by a unitary matrix $\mathcal{V}$ and has eigenvalues $E_i$. That is, $(\mathcal{V}^\dagger H \mathcal{V})_{ij} = \delta_{ij} E_i$. We define matrices $\Delta E_{ij} = E_i - E_j$, $W_{ij} = \left|\mathcal{V}_{ij}\right|^2$, and $M =W^TW$. We then have
\begin{align}
\frac{1}{n} \text{tr} \left|U(z)\right|^2  &= \frac{1}{n} \sum_{ijk} |\mathcal{V}_{ij}|^2 e^{-i(E_j - E_k) z} |\mathcal{V}_{ik}|^2\\
&=\frac{1}{n} \text{tr} \bigg[W e^{-i\Delta E z} W^T \bigg]\\
&= \frac{1}{n}\text{tr} \bigg[\cos(\Delta E\, z) M \bigg]
\end{align}
where in the last two lines the exponential and cosine are taken element-wise (i.e., are not the matrix exponential/matrix cosine) and we have used the fact that $M$ is a symmetric matrix to conclude that the corresponding sine term vanishes. Integrating this result with respect to $z$, we obtain
\begin{align}
S(z) &= \frac{1}{nz}\int_0^z dz'\, \text{tr} \left|U(z')\right|^2 \\
\label{eqn_finitez_objfunc}
 &=\frac{1}{n} \text{tr} \bigg[M \,\text{sinc}(\Delta E\, z) \bigg]
\end{align}
where the sinc is taken element-wise. This form
\begin{equation}
S(z) = \frac{1}{n} \text{tr} \bigg[M \,\text{sinc}(\Delta E\, z) \bigg]
\end{equation}
expresses $S(z)$ purely in terms of the eigenvalues and eigenvectors of the system and is useful for evaluating $S(z)$ for periodic structures like those described in the text. For aperiodic structures, it is useful to take the limit $z\to \infty$. In the absence of degeneracies, we have $\lim_{z\to \infty} \text{sinc}(\Delta E_{ij} z) = \delta_{ij}$ so that
\begin{equation}
\lim_{z\to \infty} S(z) = \frac{1}{n} \text{tr}M.
\end{equation}
Finally, noting that the diagonal entries of $M$ are equal to the eigenstate participation ratios $ P_E=\sum_i\left|\langle x_i |E\rangle \right |^4$ with $|E\rangle$ normalized, we have
\begin{equation}
\lim_{z\to \infty} S(z) = \left\langle P_E\right\rangle.
\end{equation}

\end{document}